\documentclass[a4paper,USenglish]{oasics}

\usepackage{microtype}
\usepackage{graphicx}
\usepackage{amsmath}
\usepackage{amsfonts}
\usepackage{amssymb}
\usepackage{tabularx}
\usepackage{multirow}
\usepackage{algorithm}
\usepackage{algpseudocode}
\usepackage{todonotes}
\usepackage{multicol}
\usepackage{caption}

\title{Speeding Up BigClam Implementation on SNAP}

\author[1]{C. H. Bryan Liu\footnote{Now at ASOS.com, London, UK}}
\author[1]{Benjamin Paul Chamberlain}
\affil[1]{Department of Computing, Imperial College London\\
  \texttt{liu.ch.bryan@gmail.com}
  }
\authorrunning{C.H.B. Liu and B.P. Chamberlain}

\Copyright{C.H.B. Liu and B.P. Chamberlain}

\subjclass{Information systems $\rightarrow$ Data mining; Theory of computation $\rightarrow$ Sample complexity and generalization bounds; Computing methodologies $\rightarrow$ Parallel programming languages}

\serieslogo{}
\volumeinfo
  {Billy Editor and Bill Editors}
  {2}
  {Conference/workshop/symposium title on which this volume is based on}
  {1}
  {1}
  {1}
\EventShortName{}
\DOI{10.4230/OASIcs.xxx.yyy.p}

\begin{document}

\maketitle

\begin{abstract}
We perform a detailed analysis of the C++ implementation of the Cluster Affiliation Model for Big Networks (BigClam) on the Stanford Network Analysis Project (SNAP). BigClam is a popular graph mining algorithm that is capable of finding overlapping communities in networks containing millions of nodes. Our analysis shows a key stage of the algorithm --- determining if a node belongs to a community --- dominates the runtime of the implementation, yet the computation is not parallelized. We show that by parallelizing computations across multiple threads using OpenMP we can speed up the algorithm by 5.3 times when solving large networks for communities, while preserving the integrity of the program and the result.
\end{abstract}

\section{Introduction}

Networks can represent many systems including social interactions, transport systems, financial transactions, communications infrastructure 
and biological functions. In all cases they describe interactions (edges) between dependent entities (nodes).
One of the most important and best studied fields of network science is community detection~\cite{Kuncheva2015,Newman2004c,Newman2004a}. A community can be thought as a group of nodes having a higher density of internal than external connections~\cite{fortunato2010community}. 
Early community detection algorithms partitioned small networks into disjoint regions,  assigning each node to a single community~\cite{Andersen2006b,Pothen1997}.  
Later algorithmic advances both relax the disjointness requirement (allowing overlapping communities) and scale to much larger networks. Overlapping community detection algorithms are more general than partitioning methods, which they include as special cases \cite{Evans2010a,Psorakis2011,whang2013overlapping}. Methods that focus on scaling community detection have allowed communities to be detected in networks with millions or even billions of nodes \cite{blondel2008fast,Pons2005,Xin:2013:GRD:2484425.2484427}.

\begin{figure}
  \vspace*{-20pt}
  \begin{center}
    \includegraphics[width=0.88\textwidth, trim = 3mm 0 3mm 0, clip]                    
                    {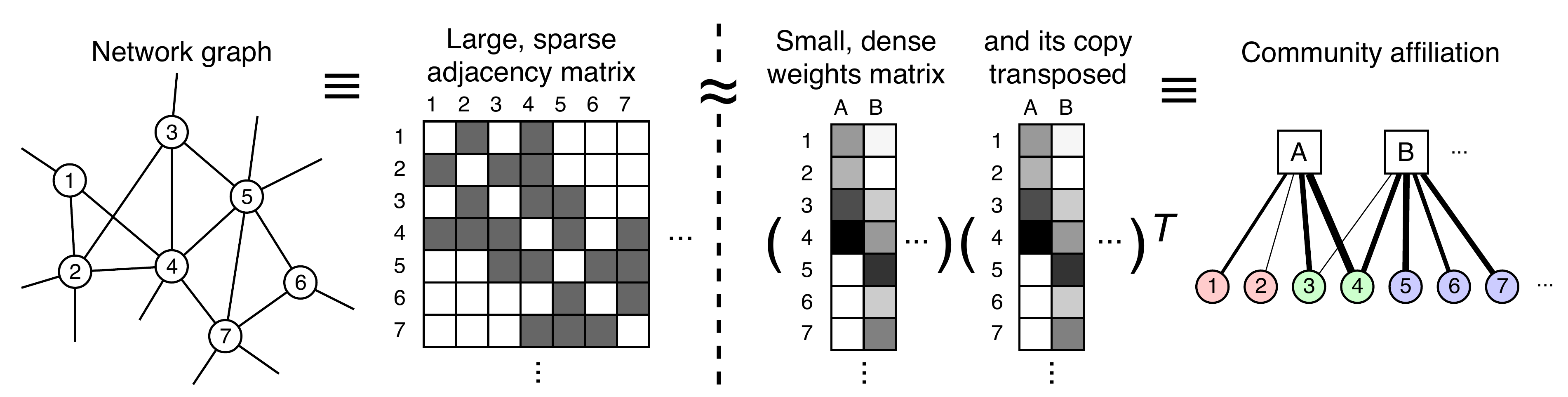}
  \end{center}
  \vspace*{-20pt}
  \caption{Illustration of community detection in a graph in terms of non-negative matrix factorization. (Left half) It is common to represent a network graph by a large, sparse adjacency matrix. (Right half) Yang and Leskovec proposed modeling community affiliations with a bipartite graph between communities and nodes, with affiliation weights represented by a small, dense, non-negative matrix~\cite{yang2013overlapping}. (Middle two panels) By finding the most likely non-negative matrix, which when multiplied with its transpose best resembles the given adjacency matrix, we can obtain the most likely community affiliation w.r.t. the given network graph.}
  \label{fig:bigclam_highlevel_illustration}
  \vspace*{-10pt}
\end{figure}

The Cluster Affiliation Model for Big Networks (BigClam), proposed by Yang and Leskovec~\cite{yang2013overlapping} is both scalable and discovers overlapping communities. 
Under BigClam, nodes can be in multiple communities, and affiliation weight between a node and a community is modeled as a positive continuous number.
The right half of Figure~\ref{fig:bigclam_highlevel_illustration} shows the affiliation weights for seven nodes to two communities.
This can be represented as a \textit{bipartite graph}, or an \textit{affiliation weights matrix}. 
The graph of a network is usually represented by a sparse binary adjacency matrix (left half).
BigClam infers the affiliation weights matrix by applying non-negative matrix factorization~\cite{hoyer2004non} to the adjacency matrix.  
The algorithm learns the affiliation weights matrix that is best able to reconstruct the underlying adjacency matrix subject to the constraints of positivity and local optimality.

BigClam is a popular and highly cited method that features in a number of lectures and tutorials~\cite{Leskovec:2014:MMD:2787930,sarkar2015lecture}. The related software project~\cite{leskovec2016snap} has attracted hundreds of GitHub stars.
Due to the popularity of this model amongst both researchers and practitioners, we perform a rigorous analysis of the C++ implementation provided on the Stanford Network Analysis Project (SNAP)~\cite{leskovec2016snap}. 
Our analysis of the BigClam source code reveals that the algorithm has three stages.
In particular, the final Community Association (CA) stage, which makes assignments of nodes to communities, generally dominates the runtime, yet its computation is not parallelized across CPU threads. The runtime domination of CA is especially true for networks with large numbers of communities, which is common in real networks (see \cite{snapnets}). Not parallelizing computation where available results in lengthened runtime and wastes available hardware resources as they are put on idle.

This motivates our work in parallelizing computation in the CA stage to speed up the BigClam implementation on SNAP. Our major consideration is the parallelization must not introduce race condition on shared objects that compromise the integrity of the results.
We parallelize the CA stage with OpenMP, a specification
for high-level parallelism in C++ programs, and we show that the parallelization 
achieves as much as 5.3 times speed up and
saves as much as 12.8 hours when
solving networks by Leskovec and Krevl~\cite{snapnets} using an eight-thread machine (Intel i7-4790 @ 3.60 GHz CPU).

To summarize, our contributions are as follow: (1) We profile the runtime of the BigClam implementation on SNAP in terms of its three stages.
(2) We show that the CA stage dominates the runtime in current BigClam implementation on SNAP when solving networks with large numbers of communities, which is common in real networks.
(3) We provide a detailed description, and the code implementation of how we parallelize computation on the CA stage, with a comprehensive discussion on avoiding race conditions.
We also provide experimental results showing that the speed up is statistically significant, and preserves the result's integrity.\footnote{All code and experiment data are available on \texttt{\url{https://github.com/liuchbryan/snap/tree/master/contrib/ICL-bigclam\_speedup}}.}


\section{SNAP Implementation: The Bottleneck}
\label{sec:existing_runtime_analysis}
We first examine the BigClam community detection algorithm and identify the bottleneck(s) in its implementation on SNAP. The core idea of BigClam is to find the affiliation weights matrix $F$ that maximizes the log-likelihood function.\footnote{The $(u, c)^{\textrm{th}}$ entry of $F$ represents the strength of the community affiliation between user~$u$ and community~$c$ in a network (see Figure \ref{fig:bigclam_highlevel_illustration} for an illustration).} The mathematical formulation is detailed in Appendix~\ref{sec:background}.

By examining its implementation on SNAP, we observe that the community detection algorithm has three stages: 
Conductance Test (CT), which initializes the affiliation strength matrix; Gradient Ascent (GA), which finds the optimal affiliation weights matrix; and Community Association (CA), which determines if an affiliation exists between a community and a node based on the value of affiliation weight recorded under the said matrix in relation to a pre-specified threshold.


We show the average-case runtime complexity of the three stages in Table~\ref{tab:runtime_complexity}. The full derivation is available in Appendix~\ref{sec:snap_runtime_complexity_analysis}. It can be seen that the CA stage will dominate the runtime if the number of communities is large, which we formalize as: $|C| \gg \frac{kr}{t^*} \left(\frac{|E|}{|V|}\right)^2$,
where $|V|$, $|E|$, $|C|$, $r$, $k$, $t^*$ represents the number of nodes, edges, communities, community affiliations per node, epochs, and the speed-up multiple achieved by parallelizing computation across threads respectively (see Appendix~\ref{sec:network_prelim}).

\begin{table}
  \vspace*{-10pt}
\begin{center}
\begin{tabular}[]{l | l | l | l}
    Stage & Conductance Test & Gradient Ascent & Community Association \\\hline
    Complexity & $O\left(|V| \left(\dfrac{|E|}{|V|}\right)^2\right)$  & $O\left(|V| \,\dfrac{kr}{t^*}\left(\dfrac{|E|}{|V|}\right)^2 \right)$ & $O(|V||C|)$
\end{tabular}
\end{center}
\vspace*{-7pt}
\caption{The average-case runtime complexity for the three stages of the BigClam community detection algorithm. $|V|$, $|E|$, $|C|$, $r$, $k$, $t^*$ represents the number of nodes, edges, communities, community affiliations per node, epochs, and the speed-up multiple achieved by parallelizing computation across threads respectively. Derivations of the complexity are detailed in 
Appendix~\ref{sec:snap_runtime_complexity_analysis}.}
\vspace*{-15pt}
\label{tab:runtime_complexity}
\end{table}

\begin{table}
\begin{center}
  \begin{tabular}[]{c | c | c | c | c } 
    & $|V|$ & $|C|$ & $|E|$ & $r$  \\ \hline
   Amazon product co-purchase network & 334,863 & 75,149 & 925,872 & 6.78 \\
   DBLP collaboration network & 317,080 & 13,477 & 1,049,866 & 2.27\\
   LiveJournal online social network & 3,997,962 & 287,512 & 34,681,189 & 1.79 \\
   Youtube online social network & 1,134,890 & 8,385 & 2,987,624 & 0.113
  \end{tabular}
\end{center}
  \vspace*{-7pt}
  \caption{Number of nodes ($|V|$), communities ($|C|$), edges ($|E|$), and the average number of affiliations ($r$) recorded in the networks by Leskovec and Krevl \protect\cite{snapnets}. }
  \vspace*{-16pt}
  \label{tab:bigclam_speedup_num_communities}
\end{table}

\begin{figure}
  \vspace*{-10pt}
  \begin{center}
    \includegraphics[width=0.85\textwidth, trim = 2mm 6mm 0mm 0mm, clip]
                    {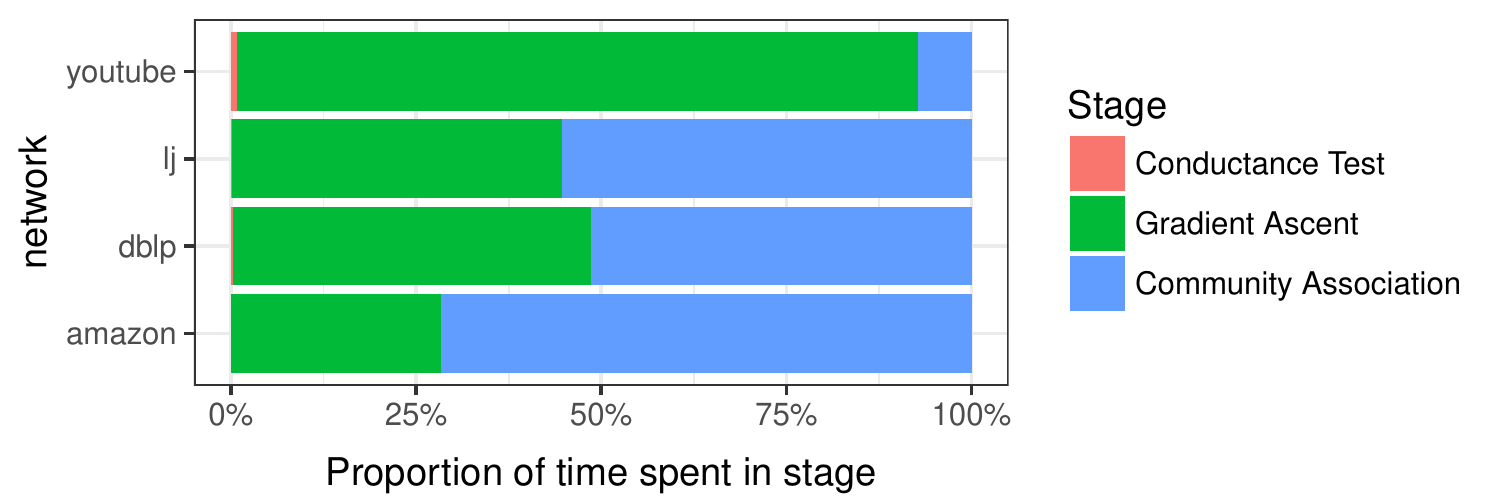}
    \vspace*{-15pt}
  \end{center}
  \caption{Average proportion of time spent on the three stages of the
           BigClam community detection algorithm 
           (From left to right: 
           Conductance Test (red), 
           Gradient Ascent (green), and 
           Community Association (blue))
           in Leskovec and Sosi\v{c}'s implementation \protect\cite{leskovec2016snap}
           for the four networks shown in Table~\ref{tab:bigclam_speedup_num_communities}.
           The implementation is tested on an eight-thread machine 
           (Intel~i7-4790 @ 3.60~GHz CPU), with the
           number of communities to detect for each network set to that recorded in Table~\ref{tab:bigclam_speedup_num_communities}.}
  \label{fig:bigclam_speedup_runtime_np_summary}
  \vspace*{-10pt}
\end{figure}

Networks satisfying the inequality above are common. For example, all
networks with ground-truth communities featured in Leskovec and Krevl~\cite{snapnets} (shown in Table~\ref{tab:bigclam_speedup_num_communities})
satisfy the inequality when $k=100$ and $t^*=4$.\footnote{A conservative estimate of the speed up achieved by parallelizing the GA stage across eight threads.} 
We confirm this by running the BigClam implementation on the networks shown in Table~\ref{tab:bigclam_speedup_num_communities} using an eight-thread machine (Intel~i7-4790 @ 3.60~GHz CPU), and measure the proportion of runtime spent in each of the three stages.
Figure~\ref{fig:bigclam_speedup_runtime_np_summary} shows the results of these experiments. While the time
spent on the CT stage is negligible, the time spent
on the CA stage generally accounts for more than half of the
entire runtime.\footnote{The Youtube network is an exception: we believe the average number of affiliations per node estimated by the BigClam algorithm is far greater than that recorded in the ground-truth (over 90\% of the nodes do not have any community affiliations). This results in a far greater value of $r$ than that reported in Table~\ref{tab:bigclam_speedup_num_communities}, which violates the inequality.}

Thus, the CA stage is usually the bottleneck in the algorithm. We notice that unlike the GA stage, in which computation is parallelized, the CA stage is not parallelized.
Therefore, the majority of the CPU resources and man-time is wasted by idling.
Parallelizing computation in the CA stage will better
utilize available resources and hence improve scalability.

\section{Speeding Up BigClam Via Parallel Computing}
\label{sec:speeding_up_bigclam}

In this section we describe how to speed up BigClam with the use of OpenMP, a specification for parallel programming \cite{openmp15} that is supported in C++ and currently used in SNAP. The goal is to speed up the CA stage while ensuring that the  input to output mapping is identical to the non-parallelized version.

\subsection{Requirements in Result Correctness}
\label{sec:req_correctness}

As with any parallel computing application, it is important to prevent race conditions between threads from undermining the correctness of the result. In the context of speeding up the CA stage of BigClam, this means that the parallelized version must produce the same set of community affiliations as the unparallelized version given the same input $F$.

\begin{figure}
  \begin{center}
    \includegraphics[width=0.85\textwidth, trim = 0 0 0 0 , clip]
                    {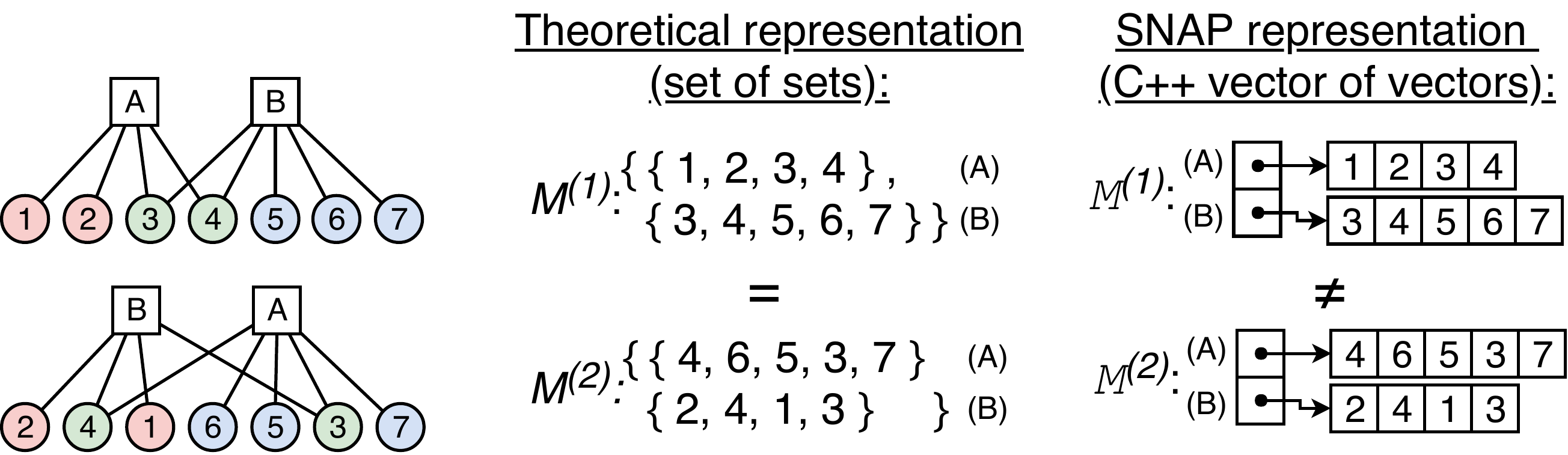}
  \end{center}
  \vspace*{-10pt}
  \caption{The two representations of the community affiliations. (Left) the underlying bipartite graph. (Middle) the theoretical representation as a set of sets. (Right) the SNAP C++ STL-like vector of vectors representation. In the theoretical representation the ordering is exchangeable, in the SNAP representation it is not.}
  \vspace*{-5pt}
  \label{fig:community_set_equality}
\end{figure}

The theoretical representation of the communities returned by the algorithm is a set of sets: $M = \{M_1, M_2, ..., M_c\}$ where each community is itself a set $M_c=\{u_1, u_2, ....u_k\}$. Two runs of BigClam produce the same output if $M^{(1)} = M^{(2)}$. 
However, the BigClam SNAP implementation uses vectors of vectors (implemented as C++ STL-like objects) instead of sets of sets and  enforces additional ordering that is not present theoretically. We denote the vector of vectors representation as $\mathcal{M}$. Using this representation it is possible to have $M^{(1)} = M^{(2)}$ and $\mathcal{M}^{(1)} \neq \mathcal{M}^{(2)}$  (see Figure~\ref{fig:community_set_equality}).

As the ordering of values in the inner vectors and the single outer vector does not matter, we can allow race conditions between threads when performing append operations into these vectors, and thus increase the degree of parallelism in our implementation.

\subsection{Methodology}
\label{sec:bigclam_speedup_methodology}

Algorithm \ref{alg:CA_stage_matrix_scan} outlines the existing CA stage implementation. The algorithm is simplified to include only operations related to scanning the matrix $F$ and extracting community affiliations.\footnote{We exclude operations such as 1) sorting the vector $\left(\sum_u F_{uc}\right)_{c\in C}$ and scan columns which have a higher total affiliation strength first, and 2) excluding communities if it does not have enough members from being included, as they are less computationally expensive than scanning the matrix.} The algorithm is rewritten in pseudocode to enhance readability.

\begin{algorithm}
\begin{multicols}{2}
\begin{algorithmic}[1]
  \State Initialize $\mathcal{M}$ as an empty vector
  \ForAll{$c \in C$} 
    \State Initialize $\mathcal{M}_c$ as an empty vector
    \ForAll{$u \in V$}
      \If{$F_{uc} \geq \delta$} 
         \State Append $u$ to $\mathcal{M}_c$ 
      \EndIf
    \EndFor
    \State Append $\mathcal{M}_c$ to $\mathcal{M}$  
  \EndFor
  \State \Return $\mathcal{M}$
\end{algorithmic}
\end{multicols}
\caption{The existing implementation of the CA stage, simplified and rewritten in pseudocode. $F$ is the affiliation weights matrix, and $\delta$ is the minimum affiliation strength threshold for a node to be considered a member of a community.}
\vspace*{-5pt}
\label{alg:CA_stage_matrix_scan}
\end{algorithm}

As discussed in Section~\ref{sec:req_correctness}, we can spread the task of scanning a particular column of~$F$ over multiple threads while maintaining correctness  --- all node IDs will be added to the correct community vector, and with the proper synchronization mechanism (see discussion below) all community vectors will be present in the final result. In the terminology of OpenMP, we can parallelize the outer \textbf{for} loop over all communities, covering operations in lines~3--8 of Algorithm~\ref{alg:CA_stage_matrix_scan}.

To prevent unintended race conditions while maintaining the highest level of parallelism, we declare a \textit{critical operation} as any operation
that involves objects that are shared between threads. Each critical operations is controlled by a mutex that prevents multiple threads from simultaneously writing to an object. It is safe to parallelize operations that involve only objects
used by a single thread (a.k.a. private objects/variables) and read-only
objects that are shared between threads.
In our case, the only object that is shared between threads and
involves write operations is the set of community affiliations $\mathcal{M}$. All other
objects are either shared and read-only, or private to a thread.

\begin{itemize}
  \item The affiliation weights matrix $F$ is read-only by all threads
  \item The lower affiliation weight threshold $\delta$ is defined as a
        C++ constant (which is unmodifiable once defined), and hence is read-only
  \item The vector / list used to keep track of current
        community's members ($\mathcal{M}_c$) is local in the scope of the outer \textbf{for} loop, and hence is private to a thread according to the OpenMP specification \cite{openmp15}.
\end{itemize}

Therefore, the only operation that needs to be declared as critical is the append to $\mathcal{M}$ in line 9 of Algorithm~\ref{alg:CA_stage_matrix_scan}. Only one thread can append $\mathcal{M}_c$ to $\mathcal{M}$ at a time while all other operations can be parallelized.

\section{Experiments}
\label{sec:experiments}

We run a number of experiments to validate the methodology described in Section~\ref{sec:speeding_up_bigclam}.
We show that parallelizing computation of the CA stage over multiple threads 1) reduces the runtime in the CA stage (and hence the overall BigClam implementation), and 2) retains the result correctness. 

We use the datasets featured in Leskovec and Krevl \cite{snapnets} (see Table \ref{tab:bigclam_speedup_num_communities} for details of the datasets), which are widely used to benchmark the runtime of overlapping community detection algorithms \cite{Wang:2015:CDS:2794367.2794370, whang2013overlapping, yang2012community, Zhang:2015:IIL:2887007.2887063}, including by BigClam itself \cite{yang2013overlapping}. 

\subsection{Runtime Reduction}
\label{sec:expt_runtime_reduction}

\begin{table*}
\vspace*{-10pt}
\begin{center}
  \begin{tabularx}{0.81\textwidth}{c|cc|cc}
    Networks & \multicolumn{2}{ c | }{CA stage} & \multicolumn{2}{ c }{Overall} \\
    & Unparallelized & Parallelized & Unparallelized & Parallelized \\ \hline
   Amazon
     & 1077.58 & 203.74 & 1505.38 & 610.23 \\
   DBLP
     & 160.39 & 30.00 & 312.47 & 180.69 \\
   LiveJournal
     & 55363.22 & 9233.02 & 100146.81 & 55259.48  \\
   Youtube
     & 213.62 & 43.07 & 2965.12 & 2717.85
  \end{tabularx}
  \vspace*{5pt}
  \caption{Average time taken, in seconds, to run \textbf{a)} the community 
           association (CA) stage \textbf{b)} the entire BigClam community 
           detection algorithm, without and with parallelization of
           the community association stage. The implementations are tested on 
           eight-thread machines with the same CPU specifications (Intel
           i7-4790 @ 3.60 GHz CPU).}
  \label{tab:bigclam_speedup_runtime_comparison}
\end{center}
\vspace*{-20pt}
\end{table*}

To demonstrate that parallelization reduces the algorithmic runtime, we run the unparallelized and parallelized variants of 
BigClam on multiple machines with Intel~i7-4790 @ 3.60~GHz CPU (eight threads) for 100 epochs.
Each machine runs only
one of the variants at any time to ensure all CPU threads are 
dedicated to one variant. 
For each run, the program detects communities in the networks specified in Table~\ref{tab:bigclam_speedup_num_communities}, with the number of communities to detect set to that specified by the table.
We measure the runtime of each stage of the BigClam for both the parallelized and unparallelized implementations across multiple runs. 

\begin{figure*}
  \vspace*{-10pt}
  \centering
    \subfloat[Amazon]{
        \includegraphics[width=0.23\textwidth,trim = 0mm 0mm 0mm 0mm, clip]{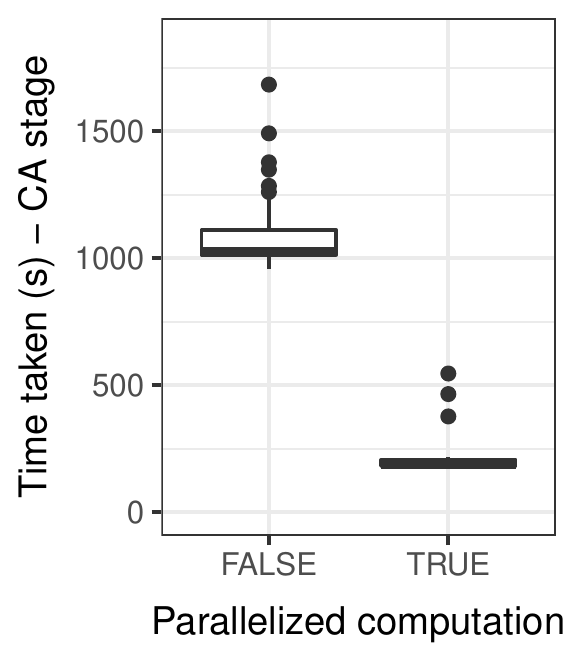}
      \label{fig:bigclam_speedup_comparison_amazon}}
    \subfloat[DBLP]{
        \includegraphics[width=0.23\textwidth,trim = 0mm 0mm 0mm 0mm, clip]{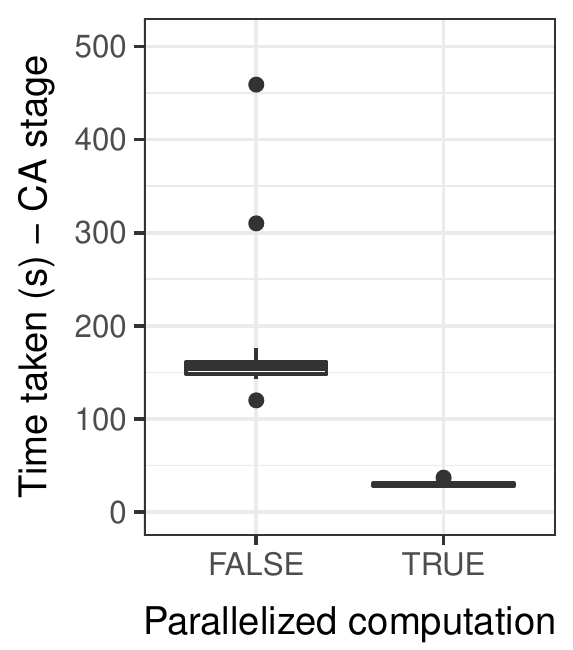}}
    \subfloat[LiveJournal]{
        \includegraphics[width=0.23\textwidth,trim = 0mm 0mm 0mm 0mm, clip]{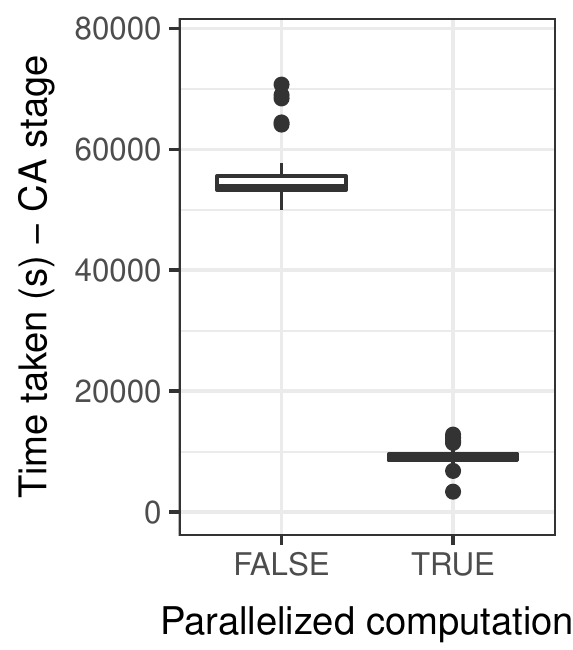}}
    \subfloat[Youtube]{
        \includegraphics[width=0.23\textwidth,trim = 0mm 0mm 0mm 0mm, clip]{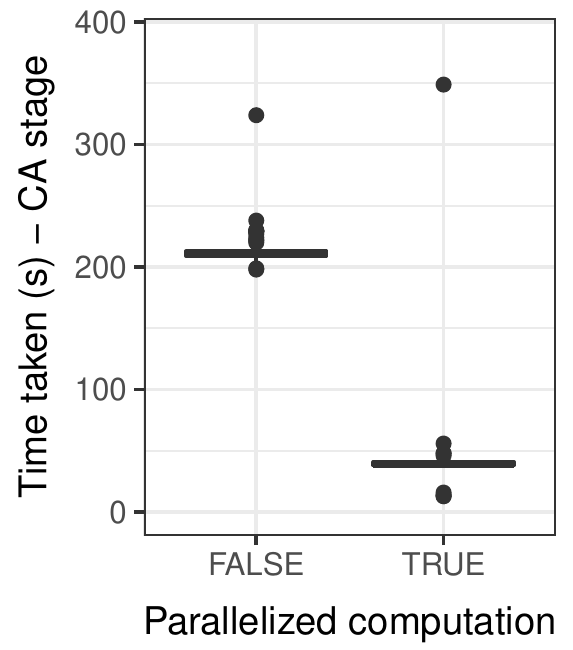}}
  \caption{Box plot of the time taken, in seconds, to run the community
           association stage on different networks with ground-truth
           communities, without and with parallelization on
           the stage. The time taken with parallelization on the stage is significantly lower.}
  \label{fig:bigclam_speedup_comparison}
\end{figure*}

The average runtime is reported in Table~\ref{tab:bigclam_speedup_runtime_comparison} and we perform a Welch's $t$-test
to determine if the parallelized implementation achieves a significantly lower runtime
for a) the CA stage, and b) the entire BigClam program.
We visualize the results of this experiment in Figure~\ref{fig:bigclam_speedup_comparison}. It is clear from Table~\ref{tab:bigclam_speedup_runtime_comparison} and Figure~\ref{fig:bigclam_speedup_comparison} that our implementation produces a significant
runtime reduction in the CA stage for all networks shown in Table~\ref{tab:bigclam_speedup_num_communities}. With an eight-thread machine, we achieve a 5.3 times speed up in the
CA stage, and subsequently a 2.5 times speed up in the overall BigClam algorithm for the Amazon product co-purchase network. For the LiveJournal network
the runtime of the CA stage is reduced by 46,130 seconds (or 12.8 hours) on average.\footnote{Yang and Leskovec state that, ``with
20 threads, it takes about one day to fit BigClam to the LiveJournal 
network''~\cite{yang2013overlapping}
--- we are able to fit this network with only eight threads in less than 16 hours.}

On the other hand, parallelizing the CA stage does not
bring massive improvements in runtime on networks with low numbers
of communities (those that do not satisfy the inequality in Section~\ref{sec:existing_runtime_analysis}). We only achieve a 1.1 times speed up on the overall runtime
solving the Youtube network, despite achieving a 4.96 times
speed up on the CA stage. The speedup is not apparent in networks where the algorithm is dominated by the GA stage, where parallelizing the computation in the CA
stage brings only marginal improvements.

\subsection{Verification of Result Correctness}

To confirm that parallelization of the CA stage produces the same set of community affiliation predictions as the unparallelized version we create a utility program. The program sorts the node IDs in a community and the communities in the program output in lexicographical
order before comparing (strict) equality. This is necessary as common approaches to compare program outputs (e.g. \texttt{diff} or MD5 check
sum) will fail even if two sets of communities are equal, as discussed in Section~\ref{sec:req_correctness}.

Our utility does not report any discrepancies between the
program outputs produced by the parallelized and unparallelized variants, hence we conclude that our parallelization in the CA stage produces the same output, backed by a theoretical discussion in Section~\ref{sec:bigclam_speedup_methodology} and experimental verification.

\section{Conclusion}
\label{sec:conclusion}

In this work we profile the runtime of the BigClam implementation on SNAP, a popular overlapping community detection algorithm on an extensively used network analysis platform. We are able to split the runtime of the algorithm into three stages --- the conductance test (initialization) stage, the gradient ascent (optimization) stage and the community association (extraction) stage --- and provide an average-case runtime complexity for each stage.

We show the community association stage is dominating the runtime in the current implementation when solving real networks, and parallelize its implementation to speed up BigClam. We show the speed up is both statistically significant and of practical utility, including a 5.3 times speed up on the community association stage (and 2.5 times overall) when solving the Amazon product co-purchase network, and saving 12.8 hours on the community association stage with an eight-thread machine. We release all relevant code and experimental data on our GitHub repository so that the research community can immediately benefit from our work and replicate our results.\footnote{\texttt{\url{https://github.com/liuchbryan/snap/tree/master/contrib/ICL-bigclam\_speedup}}}

\subparagraph*{Acknowledgements}
The authors thank Marc P. Deisenroth for useful
discussions and the anonymous reviewers for providing many improvements
to the original manuscript.

\appendix

\section{Key Formulation of BigClam Community Detection Algorithm}
\label{sec:background}
In this section we first introduce the nomenclature of networks, before moving on to the specifics of the BigClam community detection algorithm.

\subsection{Network Preliminaries}
\label{sec:network_prelim}

A network is a data structure that contains a graph and a set of attributes. A graph $G(V,E)$ is composed of a set of nodes $V$ and a set of edges $E = (v_i,v_j)$ where $v_i,v_j \in V$ that connect two nodes. The graph can be represented by an \emph{adjacency matrix} $A \in \{0,1\}^{|V|\times|V|}$ where $A_{ij}$ is one if $(v_i,v_j)\in E$ and zero otherwise. Attributes can apply to either edges or nodes. 


We denote the set of communities $C=\{c_1,c_2,...,c_n\}$, where $c_i$ indexes the $i^{th}$ community. The set of community affiliations is defined as a set of sets $M = \{M_c : c \in C\}$, where
$M_c = \{u_1, u_2, ..., u_k\}$
 is a set containing the nodes affiliated to community~$c \in C$.
 
We also denote $\mathcal{N}(u)$ as the set of neighbours of a node $u$ in $G$, and the neighborhood containing $u$ and its neighbors $N(u)$.

Part of our runtime analysis involves the average number of communities a node is affiliated with, which we formally define as:

\begin{definition}
Let $D_u = \{c: u\in M_c\}$ be the set of communities that node $u \in V$ is affiliated with.
Then the average number of community affiliations for all nodes $r$ is 
\begin{align}
  r = \frac{1}{|V|} \sum_{u \in V} |D_u| \;,
\end{align}
where $|D_u|$ is the number of communities that node $u$ is affiliated with.
\end{definition}

\subsection{BigClam Community Detection Algorithm}

The core idea of the BigClam community detection algorithm is to find the community affiliation weights matrix $F=(F_{uc})_{u \in V, c \in C}$, where the $(u, c)^{\textrm{th}}$ entry represents the strength of the community affiliation between user $u$ and community $c$ in a network (see Figure \ref{fig:bigclam_highlevel_illustration} for an illustration), that maximizes the log-likelihood function.
Yang and Leskovec~\cite{yang2013overlapping} use an iterative approach, where at each iteration they fix the affiliation weights for all but one node (say $u$), and perform a gradient ascent on the affiliation weights for node $u$.
The log-likelihood for the corresponding row $\vec{F_u} = (F_{uc})_{c\in C}$ of $F$ is specified as:
\begin{align}
  l(\vec{F_u}) = \sum_{v \in \mathcal{N}(u)} \log\left(1 - \exp(-\vec{F_u}{\vec{F_v}}^T)\right)
           - \sum_{v \not\in \mathcal{N}(u)} \vec{F_u}{\vec{F_v}}^T \;.
\label{eq:bigclam_row_log_likelihood}
\end{align}
We follow the original BigClam notation and so $\vec{F_u} {\vec{F_v}}^T$ is an \emph{inner product}.

Differentiating Equation \eqref{eq:bigclam_row_log_likelihood} w.r.t. $\vec{F_u}$ gives the gradient:
\begin{align}
  \triangledown l(\vec{F_u}) &=
  \sum_{v \in \mathcal{N}(u)} F_v 
    \frac{\exp(-\vec{F_u} {\vec{F_v}}^T)}{1- \exp(-\vec{F_u} {\vec{F_v}}^T)}
  - \sum_{v \not\in \mathcal{N}(u)} \vec{F_v} 
  \label{eq:bigclam_log_likelihood_gradient} \\
  & = \sum_{v \in \mathcal{N}(u)} F_v 
    \frac{\exp(-\vec{F_u} {\vec{F_v}}^T)}{1- \exp(-\vec{F_u} {\vec{F_v}}^T)}
  - \left(\sum_{v \in V} \vec{F_v} - \vec{F_u} 
    - \sum_{v \in \mathcal{N}(u)} \vec{F_v}\right)\;.
  \label{eq:bigclam_gradient_optimisation}
\end{align}
In Equation~\eqref{eq:bigclam_gradient_optimisation} $\sum_{v \in V} \vec{F_v} $ can be precomputed, and $\sum_{v \in \mathcal{N}(u)} \vec{F_v}$ is computed on each gradient evaluation. This results in a more computationally efficient formulation as network graphs are usually sparse (i.e. $|\mathcal{N}(u)| \ll |V|$).

The BigClam community detection algorithm initializes $F$ as:
\begin{align}
  F_{(u')(N(u))} =
  \left\{
  \begin{array}{ll}
    1 \quad & \textrm{if $u' \in N(u)$ and $N(u)$ is a locally}\\
         & \textrm{minimal neighborhood~\cite{Gleich:2012:VNL:2339530.2339628} of $u$}\\
    0    & \textrm{otherwise}  
  \end{array}
  \right. \;,
\end{align}
where $N(u)$ represents $u$ and its neighbours in $G$, and regards $u \in V$ as a member of 
$c \in C$ from the most likely affiliation weights matrix $F$ if:
\begin{align}
F_{uc} \geq \delta = \sqrt{-\log(1-\epsilon)} \;, 
\label{eq:community_affiliation_criterion}
\end{align}
where $\epsilon = \frac{2|E|}{|V|(|V|-1)}$ is the
background probability for a random edge to form in the graph.

\section{SNAP Implementation: A Runtime Complexity Analysis}
\label{sec:snap_runtime_complexity_analysis}
We observe the BigClam community detection algorithm has three stages: Conductance Test, Gradient Ascent, and Community Association. Here we derive the runtime complexity for each of the three stages. 

\subsection{The Conductance Test Stage} 
\label{sec:CT_stage_complexity}


The algorithm begins by testing each node to see if it belongs to a locally minimal neighborhood as defined by Gleich et al.~\cite{Gleich:2012:VNL:2339530.2339628}. 
The initial / seed communities are chosen to be the locally minimal neighborhoods.

For each node $u \in V$ we calculate the conductance of its neighborhood. The conductance of an neighbourhood $N(u)$ is the fraction of 
edges from nodes within $N(u)$ to nodes in the same neighborhood over that to nodes outside the neighborhood~\cite{Kannan:2004:CGB:990308.990313}.
This involves traversing each neighbor $v\in \mathcal{N}(u)$ and finding out how many members of $\mathcal{N}(v)$ are not in $N(u)$. Hence there are
$\sum_{u \in V}  \sum_{v \in \mathcal{N}(u)} |\mathcal{N}(v)|$ operations involved.

We simplify 
the expression above
by replacing $|\mathcal{N}(u)|$  $\forall u \in V$ with the average number of neighbors, and using the fact that it is by definition the average degree of the network graph ($|E|/|V|$). This leads to an average-case complexity of 
$O\left(|V| \frac{|E|}{|V|} \frac{|E|}{|V|}\right)$.

\subsection{Gradient Ascent Stage}

After initialization, the algorithm optimizes the affiliation weights matrix $F$ to maximize the log-likelihood function (see Equation~\eqref{eq:bigclam_row_log_likelihood}) using gradient ascent. To understand the runtime complexity of the GA stage, we first look at the two building blocks --- calculating the dot product and summing $\vec{F_v}$ --- and their runtime complexity in the implementation.

\subsubsection{Dot Product Runtime Complexity}
Calculating the dot product between two vectors $\vec{F_u}$ and $\vec{F_v}$ is required to calculate the gradient given in Equation~\eqref{eq:bigclam_log_likelihood_gradient}. This is performed on each pairs of connected nodes in $G$ for each epoch.

In a na\"{i}ve implementation that sums the product of the corresponding (dense) vector elements, the number of operations required scales with the length of $\vec{F_u}$.  Many such operations in this context are unnecessary --- a node $u$ in real networks is likely to be affiliated with only a small number of communities,\footnote{Liu \cite{liu2016thesis} has shown the the maximum number of affiliations for any node is 116 out of 75,149 possible communities in the Amazon product co-purchase network, and 682 out of 957,154 possible communities in the LiveJournal social network.} leading to a large number of entries in $\vec{F_u}$ being set to zero (as $u$ is unaffiliated to those communities).

The SNAP implementation stores $\vec{F_u}$ as a sparse vector, where only non-zero elements are recorded along with its position. Using sparse vectors, the number of operations for a dot product between $F_u$ and $F_v$ scales as:
\begin{align}
   d_{\textrm{DP}}(u, v) \triangleq \min\left(| D_u |, | D_v |\right),
\label{eq:num_op_sparse_vector}
\end{align}
which is the minimum number of community affiliations possessed by the two nodes $u$ and $v$.

\subsubsection{Vector Sum Runtime Complexity}
We then consider the number of elements to be traversed in each $\vec{F_v}$ when calculating the sum of affiliation weights for all neighbors of a node $u$, $\sum_{v \in \mathcal{N}(u)} \vec{F_v}$. The sum is featured in Equation \eqref{eq:bigclam_gradient_optimisation} as part of the gradient calculation. Similar to the dot product calculation, implementing $\vec{F_v}$ as sparse vectors means the algorithm need not consider all $|C|$ affiliation weights in $\vec{F_v}$ but only the weights associated with communities that the neighbor nodes are a member of:
$\bigcup_{v \in \mathcal{N}(u)} D_v  \;\subseteq  C$ .

The cardinality of the set in
the expression above is bounded above by
\begin{align}
    d_{\textrm{VS}}(u) \triangleq |\mathcal{N}(u)| \max_{v \in \mathcal{N}(u)}(| D_v | ),
    \label{eq:vector_sum_comms_UB}
\end{align}
assuming all neighbors of node $u$ belong to disjoint sets of communities.\footnote{In practice the cardinality will be much smaller due to the ``small world" phenomenon: a node's neighbors are likely to be connected themselves \cite{watts1998collective}, which according to BigClam is due to them being mutual members of one or more communities.}

\subsubsection{Overall Runtime Complexity of the GA Stage}
We can now estimate the runtime complexity of the GA stage. In this stage the algorithm iterates over the nodes multiple times, calculates the row gradient in Equation~\eqref{eq:bigclam_log_likelihood_gradient} and updates the affiliation weights. This is done until the convergence criteria is met, or for a pre-specified number of epochs.

Equation~\eqref{eq:bigclam_log_likelihood_gradient} shows that the gradient of the log-likelihood is the difference of two summations. The first summation involves calculating the vector sum and dot product over each neighbor $v\in\mathcal{N}(u)$, and the second summation involves calculating the vector sum over $v\notin \mathcal{N}(u)$. This is made more efficient by Equation~\eqref{eq:bigclam_gradient_optimisation} as real graphs are sparse and so the number of neighbors of a node is far less than the number of non-neighbors. The number of operations required in calculating the row gradient is then bounded above by:
\begin{align}
    \gamma \left[\sum_{v \in \mathcal{N}(u)} \left(d_{\textrm{VS}}(u) + d_{\textrm{DP}}(u, v)\right) + \sum_{v \in \mathcal{N}(u)} d_{\textrm{VS}}(u) \right],
    \label{eq:ga_operations_one_row}
\end{align}
where $\gamma$ is a constant multiplier. 

We notice that $d_{\textrm{VS}}(u)$ dominates $d_{\textrm{DP}}(u, v)$ $\forall v \in \mathcal{N}(u)$, and hence Expression~\eqref{eq:ga_operations_one_row} can be further simplified to
$\gamma' \left[ |\mathcal{N}(u)| \, d_{\textrm{VS}}(u) \right]$,
where $\gamma'$ is another constant multiplier.

We replace $|\mathcal{N}(u)|$ by $|E|/|V|$, just as we did in Section~\ref{sec:CT_stage_complexity}. Furthermore we approximate 
$d_{\textrm{VS}}(u)$ (see Equation~\eqref{eq:vector_sum_comms_UB})
in the average case by $|E|/|V| \times r$. 

We have to calculate the row gradient for all $|V|$ nodes over $k$ epochs (which we specify). Moreover, the computation of the GA stage is parallelized onto $t$ threads using OpenMP~\cite{openmp15}, which will reduce the runtime by $t^* \leq t$ folds due to synchronization overhead~\cite{Wu:2017:EEI:3067421.3067427}. We arrive at our average-case complexity of 
\begin{align}
O\left(\frac{k}{t^*} |V|\frac{|E|}{|V|}\frac{|E|}{|V|} r\right).
  \label{eq:GA_runtime_complexity}
\end{align}

Note that $r$ in Expression~\eqref{eq:GA_runtime_complexity} is the BigClam estimate of the average number of affiliations per node, not the value realized in the ground-truth, and the two values can differ significantly.

\subsection{Community Association Stage}
\label{sec:CA_stage_complexity}
The final stage of the algorithm takes the most likely community affiliation weights matrix~$F$, and for each community $c \in C$ and each node $u \in V$ determines if $u$ is affiliated to $c$ by examining the entry $F_{uc}$ (see Equation \eqref{eq:community_affiliation_criterion}). 

The implementation treats rows of $F$ as dense vectors, and requires scanning through all entries of $F$ to determine all community affiliations. Hence, at least $|C||V|$ comparisons must be performed, leading to an average-case complexity of $O\left(|C||V|\right)$.

\bibliography{references}

\end{document}